\documentclass[
aps,
twocolumn,
amsmath,
amssymb,
floatfix,
showpacs,
superscriptaddress,
nofootinbib,
longbibliography
]{revtex4-2}
\usepackage[T1]{fontenc}
\usepackage{mathtools}
\usepackage{braket}
\usepackage{bm}
\usepackage{MnSymbol}
\usepackage{booktabs}
\usepackage{graphicx}
\usepackage{tikz}
\usepackage{pgffor}
\usepackage{blkarray}
\usepackage{sidecap}

\usepackage[dvipsnames]{xcolor}
\usepackage{enumitem}
\usepackage[normalem]{ulem}
\usepackage{silence}
\usepackage[
colorlinks,
linkcolor=NavyBlue,
citecolor=NavyBlue,
urlcolor=NavyBlue
]{hyperref}

\mathchardef\mhyphen="2D

\newcommand\bea{\begin{eqnarray}}
\newcommand\eea{\end{eqnarray}}
\newcommand\beq{\begin{equation}}
\newcommand\eeq{\end{equation}}

\definecolor{lime}{HTML}{A6CE39}

\DeclareRobustCommand{\orcidicon}{%
\hspace{-1.0mm}
\begin{tikzpicture}
  \draw[lime, fill=lime] (0,0)
  circle [radius=0.15]
  node[white] {{\fontfamily{qag}\selectfont \tiny \,ID}};
  \draw[white, fill=white] (-0.0525,0.095)
  circle [radius=0.007];
\end{tikzpicture}
\hspace{-3.0mm}
}

\foreach \x in {A,...,Z}{%
  \expandafter\xdef\csname orcid\x\endcsname{%
    \noexpand\href{https://orcid.org/\csname orcidauthor\x\endcsname}%
    {\noexpand\orcidicon}}%
}

\begin{document}

\title{Electron–phonon control of obstructed and quantum spin Hall topology in an altermagnet}
\author{Kuntal Bhattacharyya\orcidA{}}
\email[]{kuntalbhat22@gmail.com} 
\affiliation{Department of Physics, Indian Institute of Technology Guwahati, Guwahati-781039, Assam, India}
\author{Bilal Tanatar\orcidB{}}
\email[]{tanatar@fen.bilkent.edu.tr}
\affiliation{Department of Physics, Bilkent University, 06800 Bilkent, Ankara, Türkiye}
\author{Saurabh Basu}
\email[]{saurabh@iitg.ac.in}
\affiliation{Department of Physics, Indian Institute of Technology Guwahati, Guwahati-781039, Assam, India}
\begin{abstract} 
We theoretically demonstrate control over the interplay between obstructed atomic insulating and first-order topological phases via electron-phonon coupling (EPC) in a \(d\)-wave altermagnet (AM) that hosts nonrelativistic spin-split bands with zero net magnetization. By embedding a $d_{x^{2}-y^{2}}$-wave AM in a quantum spin Hall (QSH) insulator, we show that an altermagnetic order beyond a critical strength reconstructs the parent QSH state of a Bernevig-Hughes-Zhang (BHZ)-type topological insulator (TI) into an obstructed atomic insulator (OAI), characterized by nontrivial bulk polarization with quantized fractional corner charges, accompanied by characteristic in-gap boundary states, while the spin Chern number vanishes. Upon incorporating sufficiently strong EPC, the polaronic dressing reorganizes the Wannier centers and the associated effective edge topology, thereby weakening the altermagnetism-induced OAI phase and restoring the helical QSH edge modes. Further, EPC eventually drives the system into a trivial insulator. By uncovering a controllable switching mechanism between topologically distinct insulating phases in AMs, our study establishes a direct route toward phonon-tuned spintronic functionalities in unconventional magnetic platforms.    
\end{abstract}
\maketitle
\noindent\textit{\textcolor{blue}{Introduction:-}}
Over the last two decades, TIs have reshaped modern condensed matter physics by revealing how the nontrivial insulating bulk can coexist with robust conducting boundary states characterized by distinct topological invariants~\cite{Thouless1982, Haldane1988, Hasan2010, Qi2011, Nagaosa2010, Kane2005, Bernevig2006}. In contrast to the helical edges of a two-dimensional (2D) QSH insulator ($\mathbb{Z}_2$-insulators) described by the Kane-Mele~\cite{Kane2005} and BHZ models~\cite{Bernevig2006}, the higher-order TIs (HOTIs) exhibit corner or hinge modes~\cite{Fu2011,Ando2015,Benalcazar2017,Schindler2018,BenalcazarII2017,Schindler2018,SchindlerII2018,Ezawa2018,Benalcazar2019,Ren2020, Chen2020,Wang2022,Ma2023,Lahiri2024}. In the language of topological quantum chemistry~\cite{xu2021, xuII2021,Elcoro2021}, certain HOTI phases can arise as OAIs, where symmetry-pinned Wannier charge centers (WCCs) are displaced from the occupied atomic Wyckoff positions~\cite{Soluyanov2011, Taherinejad2014,Gresch2017, Bradlyn2017, Cano2018, xu2021, xuII2021,Lee2020,Wang2022, Ma2023}. When a 2D OAI is restricted to a finite geometry, this mismatch yields in-gap corner-localized states, providing a real-space manifestation of a HOTI phase~\cite{Lee2020,xu2021}. Although these HOTI phases have been realized in artificial acoustic~\cite{Ma2019,Ni2019,Qi2020} and photonic~\cite{Serra-Garcia2018, Mittal2019, ElHassan2019} setups the electronic realization requires suitable materials and stringent symmetry (e.g., inversion, time-reversal or crystalline symmetries) controlled protocols~\cite{Ren2020, Chen2020}, involving subtle spin, orbital, and lattice intricacies. \\
Of late, an unequivocal upsurge in research has been observed on a newly discovered unconventional magnetic system, dubbed \textit{\text{``altermagnet''}}, which demonstrates a distinct class of compensated collinear magnetic order~\cite{Libor2022,LiborII2022,Liu_AM2022,Mazin2023,Betancourt2023,Bai2024,Hariki2024,Krempasky2024,LeeII_AM2024, Chakraborty_A2024, Lange2026}. In contrast to conventional magnets, namely ferromagnets (exhibiting a net magnetization) and antiferromagnets (hosting spin-degenerate bands throughout the entire Brillouin zone), AMs combine spin-orbit coupling (SOC)-free, anisotropically momentum-dependent spin splitting with zero net magnetization~\cite{Hayami2019,Yuan2021,Liu_AM2022,Libor2022,LiborII2022,Krempasky2024}. Depending on the underlying nontrivial spin-group symmetry, AMs can be classified into distinct even-parity classes, namely $d$-, $g$-, and $i$-waves~\cite{LiborII2022,Lange2026}, which entails unification of several spin-dependent phenomena, such as $d$-wave superconducting pairing potential~\cite{LiborII2022, Zhang2024, Maeda2025,Fukaya2026}, Fermi-liquid instabilities~\cite{Ahn2019}, anomalous Hall effects~\cite{Libor_AHE2020,Feng2022,Betancourt2023,Reichlova2024}, giant magnetoresistance~\cite{Shao2021,Libor_GMR2022}, spin polarization~\cite{Debnath2026,Hoi2026}, topological superconductivity~\cite{Zhu2023, Sun2023, Papaj2023, Li_AM_SC2023, Ghorashi2024,  Zhang2024,Li-AM2024, Nagae2025, Liudeng2026, Yang2026, Mcardle2026, Fukaya2026}, finite-momentum Cooper pairing~\cite{Zhang_cooper2024,Chakraborty2024,Maeda2025}, etc. Recent experiments have verified some of these features in real materials, namely MnTe~\cite{Betancourt2023,Mazin2023,LeeII_AM2024,Hariki2024,Krempasky2024}, CrSb~\cite{Reimers2024,Zeng2024,Lu2025}, and $\text{Mn}_5\text{Si}_3$~\cite{ Reichlova2024,Miina2024,Rial2024,Badura2025}. The hallmark signature of nonrelativistic spin-split altermagnetic bands connecting two magnetic sublattices with opposite spin orientations is enforced by time-reversal symmetry (TRS) breaking together with specific crystalline symmetries (such as rotations), thus hinting at alternative routes to realize and control distinct topological phases with discernible edge-state features characterized by suitable topological invariants~\cite{Ezawa2024, Rafael2025, Li-AM2025,Liu-AM2025,Liu-AM2026,Huo2026,Zhang2026,Ning2026,Ji2026}. Recent studies on AMs have triggered a renewed interest for realizing HOTI states in bilayer systems~\cite{Liu-AM2025,Liu-AM2026}, Lieb lattice~\cite{Huo2026}, $\text{Fe}_2\text{Se}_2\text{O}$ monolayer~\cite{Zhang2026}, chromium-based group-IV chalcogenides~\cite{Ning2026}, and other heterostructures~\cite{Ezawa2024,Ji2026}. However, the switchover between first- and higher-order topological states in AMs and the control of phase transitions, especially via interactions, remain underexplored.\\
EPC has recently been tested as an alternative and efficient tool to manipulate band topology~\cite{Calvo2018, Cangemi2019,Camacho2019,Sergi2020,Pimenov2021,Hu2021, Medina2022,Lu2023, Islam2024,Bhattacharyya2024, Lahiri2025, Debnath2025} by exploiting the interaction between mobile charge carriers and the coherent phonons, which can produce “dressed” quasiparticles, called polarons~\cite{Alexandrov2010}. In particular, polarons have unveiled distinct topological features in several low-dimensional platforms, such as hexagonal lattices~\cite{Calvo2018,Cangemi2019,Camacho2019,Pimenov2021,Lu2023,Islam2024,Bhattacharyya2024, Zhang2025}, square lattices~\cite{Lahiri2025}, 
chiral phonons~\cite{Medina2022,Zhu2018,Zhang2025}, Mott insulators~\cite{Sergi2020}, etc., which
add further merit to EPC-based topological engineering in 
quantum materials. Notwithstanding, the inherent complexities of SOC-free spin-split bands, nontrivial symmetries, and instabilities~\cite{Heinsdorf2025} of AMs result in limited exposure to polaronic interactions in compensated magnet-based topological platforms. Although a few propositions have been advocated to explore interaction-driven emergent phenomena in AMs~\cite{Leeb2024, Chakraborty2024,Re2025,Leraand2025,Iorsh2025,He2025,  Giuli2025,Park2026,Hoi2026}, the role of lattice excitations in tuning AM-based OAI and QSH phases has not been established yet, to the best of our knowledge.
\\To this end, the following pertinent questions arise: can altermagnetism stabilize an OAI phase with fractional corner charges and a bulk topological characterization distinct from that of the parent QSH state? If yes, how does lattice dressing compete with the altermagnetic band reconstruction, and can their interplay drive transitions among obstructed, first-order topological, and trivial insulating phases?  In this work, we address these questions by first elucidating a  $d_{x^{2}-y^{2}}$-wave AM-driven reconstruction of the parent BHZ-QSH phase and demonstrating the emergence of an OAI regime characterized by quantized bulk polarization and fractional corner charge \(e/4\) corresponding to the in-gap boundary states. We further incorporate Holstein-type EPC~\cite{Alexandrov2010,Cangemi2019, Lu2023, Islam2024,Bhattacharyya2024,Lahiri2025} and show that coherent phonon dressing reasonably suppresses the altermagnetic attributes and reorganizes the WCCs and boundary termination to restore the helical QSH edges, thereby facilitating a phononic-knob control of distinct topological phases in 2D AMs.\\ 
\textit{\textcolor{blue}{Altermagnetic TI model:-}}
\begin{figure}
    \centering
    \includegraphics[width=1\linewidth]{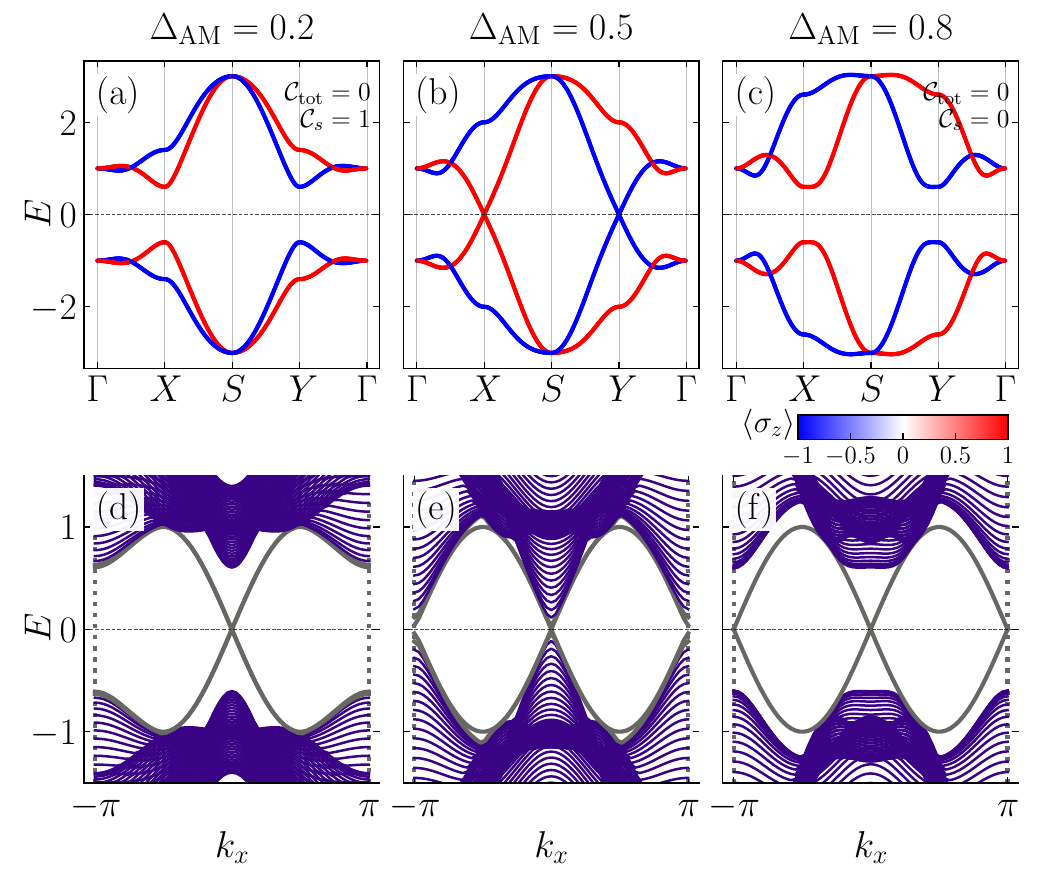}
    \caption{(a)--(c) Evolution of the bulk spectra of the AM-BHZ model with increasing altermagnetic strength $\Delta_{\mathrm{AM}}$, displaying that the
 bulk spectrum undergoes a gap-closing transition at the critical value $\Delta_{\mathrm{AM}}^c=0.5$, distinguishing the spin-Chern ($\mathcal{C}_s$) phase from the topologically distinct regime at larger $\Delta_{\mathrm{AM}}$.
(d)--(f) corresponding edge spectra, showing the progressive reconstruction of the boundary states across the transition, with panel (f) displaying the SEB characteristics of the $\Delta_{\mathrm{AM}}>0.5$ regime.}
    \label{fig:bulk plot}
\end{figure}
The collinear N\'eel ordering parametrized by
$\hat{\mathbf n}=(\sin\theta,0,\cos\theta)$, makes a $d_{x^{2}-y^{2}}$-wave AM texture take the form in the $xy$ plane as~\cite{Hayami2019,Libor2022,Li-AM2024}
\begin{equation}
\mathcal{H}_{\rm AM}(\mathbf{k},\theta)
=
\Delta_{\rm AM}(\cos k_x-\cos k_y)
\left(\sin\theta\,\sigma_x+\cos\theta\,\sigma_z\right)\otimes\tau_z .
\end{equation}
Here, $\Delta_{\text{AM}}$ is the altermagnetic strength and $\theta$ specifies the orientation of the N\'eel vector in spin space, while the factor $f_{\rm AM} (\mathbf{k})=(\cos k_x-\cos k_y)$ fixes the $d_{x^{2}-y^{2}}$ crystalline character of the spin splitting for the choice $\theta=0$ adopted here, with $\sigma$ ($\tau$) denoting the spin (orbital) indices. To address a compensated altermagnetic order-controlled QSH phase we embed a $d_{x^2-y^2}$-wave AM texture on a $4\times4$ BHZ TI model~\cite{Bernevig2006}, described by the following states in $\sigma$ and $\tau$ bases ($|s,\uparrow\rangle$, $|s,\downarrow\rangle$, $|p_x + ip_y,\uparrow\rangle$, and $|p_x-ip_y,\downarrow\rangle$) at each site of a square lattice. The composite AM-BHZ Hamiltonian under completely periodic boundary conditions is expressed in the $k$-space as~\cite{Bernevig2006,Hayami2019,Libor2022,Li-AM2024}
\setlength\abovedisplayskip{3pt}%
\setlength\belowdisplayskip{3pt}%
\begin{align}
\mathcal{H}_{\text{AM}}^{\text{BHZ}}(k) = &\left[\mathcal{M}-t(\cos k_x + \cos k_y)\right]\sigma_0\otimes\tau_z + \nonumber \\ &t_{sp}\left[\sin k_x(\sigma_z\otimes\tau_x)+\sin k_y(\sigma_0\otimes\tau_y)\right]\nonumber\\&+\Delta_{\text{AM}}(\cos k_x - \cos k_y)\sigma_z\otimes\tau_z.
\label{Ham:AM BHZ}
\end{align}
Here, $\mathcal{M}$, $t$, and $t_{sp}$ denote usual real-valued BHZ parameters~\cite{Bernevig2006}. A four-fold rotation $C_{4z}$ transforms the $d$-wave form factor as $f_{\rm AM}(C_{4z}\mathbf{k})=-f_{\rm AM}(\mathbf{k})$, whereas the TRS ($\mathcal{T}$) flips the spin polarization as $\mathcal{T}\sigma_z\mathcal{T}^{-1}=-\sigma_z$. Hence, the combined antiunitary operation $C_{4z}\mathcal{T}$ remains a symmetry, thereby relating the two spin sectors while preserving the characteristic zero-net-magnetization spin splitting~\cite{Libor2022,LiborII2022}. The AM-BHZ Hamiltonian~\eqref{Ham:AM BHZ} also conserves $s_z$, which yields two independent diagonal $2\times2$ spin blocks~\cite{Rafael2025} related through $\mathcal{C}_4\mathcal{T}$ symmetry as
$\mathcal{H}_{\uparrow}(\mathbf{k})
=
\mathcal{U}_{\mathcal{C}_4}
\mathcal{H}_{\downarrow}^{*}(-\mathcal{C}_4\mathbf{k})
\mathcal{U}_{\mathcal{C}_4}^{\dagger}$
(see details in Supplemental Material (SM)).\\
We begin by referring to Fig.~\ref{fig:bulk plot}, which clearly shows the altermagnetic spin splitting along the high-symmetry $k$- path: $(0,0)\rightarrow(\pi,0)\rightarrow(\pi,\pi)\rightarrow(0,\pi)\rightarrow(0,0)$, symbolized $\Gamma\rightarrow X\rightarrow S\rightarrow Y\rightarrow\Gamma$, in both the occupied and unoccupied subspaces of a $4$-band AM-BHZ model, as measured by its spin-polarization expectation value, $\langle\sigma_z\rangle$. We fix our model parameters as $\mathcal{M}=t=t_{sp}=1$ throughout. To ascertain the topological nature of the bulk gap that exists for $\Delta_{\text{AM}}=0.2$ and $\Delta_{\text{AM}}=0.8$ as illustrated in Fig.~\ref{fig:bulk plot}$(a)$ and $(c)$, respectively, we compute the spin Chern number as $C_s=\frac{1}{2}(\mathcal{C}_{\uparrow}^--\mathcal{C}_{\downarrow}^-)$ for the occupied subspace (labeled with `$-$' and defined as $\mathcal{H}_{\text{occ}}=\{E_{\uparrow}^-,E_{\downarrow}^-\}$), with $\mathcal{C}_{\uparrow/\downarrow}^-$ (the spin-resolved occupied projectors being well-defined) corresponding to the $\mathcal{H}_\uparrow$ and $\mathcal{H}_\downarrow$ blocks being expressed in terms of the respective spin Berry curvatures~\cite{Avron1988,Fukui2005}, namely $\Omega_{\uparrow/\downarrow}^-(k)=i[\langle\partial_{k_x}u_{\uparrow}^-|\partial_{k_y}u_{\uparrow}^-\rangle-\langle\partial_{k_y}u_{\uparrow}^-|\partial_{k_x}u_{\uparrow}^-\rangle]$ as $\mathcal{C}_{\uparrow/\downarrow}^-=\frac{1}{2\pi}\int_{BZ}d^2k~\Omega_{\uparrow/\downarrow}^-(k)$. Our calculation unfolds $\mathcal{C}_{\uparrow/\downarrow}^-=\pm1$ for $\Delta_{\text{AM}}=0.2$ (Fig.~\ref{fig:bulk plot}$(a)$), making  $\mathcal{C}_s=1$, while the vanishing of the total Chern number ($\mathcal{C}_{\text{tot}}=\mathcal{C}_{\uparrow}^-+\mathcal{C}_{\downarrow}^-$) corroborates a clear signature of QSH-type TI. However, for $\Delta_{\text{AM}}=0.8$ (Fig.~\ref{fig:bulk plot}$(c)$), $\mathcal{C}_{\uparrow/\downarrow,-}$ vanishes, suggesting a non-QSH trivial insulator in the language of a first-order topology, accompanied by a band-gap closing transition at $\Delta_{\text{AM}}=\Delta_\text{AM}^c=0.5$ (Fig.~\ref{fig:bulk plot}$(b)$). The edge bands computed on a semi-infinite ribbon geometry (Figs.\ref{fig:bulk plot}$(d)-(f)$) reveal that the AM-BHZ model retains its QSH character, as understood from the helical edge states (Fig.\ref{fig:bulk plot}$(d)$) for $0\leq\Delta_{\text{AM}}<\Delta_\text{AM}^c$. Intriguingly, beyond the bulk-gap transition point ($\Delta_\text{AM}^c$) for any $\Delta_{\text{AM}}\in[0.5:1.0]$, in-gap states continue to exist (Fig.\ref{fig:bulk plot} $(f)$), albeit with counterpropagating edge states that cancel each other at each ribbon boundary. Hence, these modes, referred to as in-gap \textit{\text{``suspended''}} edge bands (SEBs) can be easily eliminated by an adiabatic deformation of the parent Hamiltonian \eqref{Ham:AM BHZ} and thus carry no QSH signatures (satisfying $\mathcal{C}_s=0$), which require closer inspection to unfold their \textit{\text{``true''}} topological identity.
\begin{figure}
    \centering
    \includegraphics[width=1\linewidth]{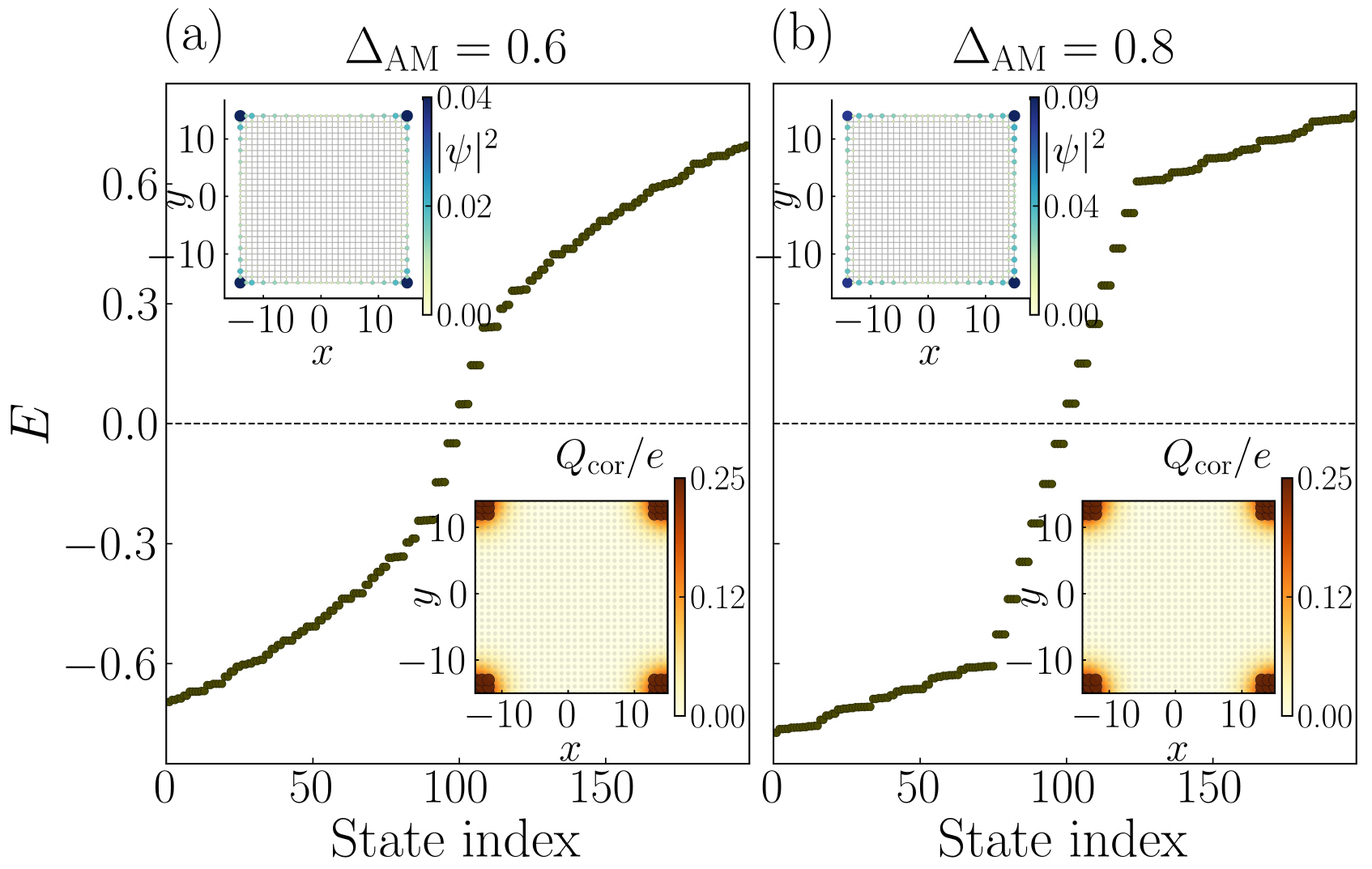}
    \caption{Open-boundary energy spectra for $(a)$ $\Delta_{\mathrm{AM}}=0.6$ and $(b)$ $\Delta_{\mathrm{AM}}=0.8$, corresponding to the parameter regime where the SEBs emerge. The states closest to zero energy appear as isolated in-gap modes, and upper insets (lower insets) denote their corner-localized probability density $|\psi|^2$  (quantized fractional corner charge  $Q_{\mathrm{cor}}\simeq\frac{e}{4}$ accumulation), indicating an AM-OAI phase.}
    \label{fig:corner plot}
\end{figure}\\
\textit{\textcolor{blue}{Corner modes, charge fractionalization, and topological characterization:-}} To inspect the fate of these first-order topologically trivial SEBs for the $\Delta_\text{AM}>\Delta_\text{AM}^c$ regime, we compute the real-space bandstructures of the AM-BHZ Hamiltonian~\eqref{Ham:AM BHZ} under open-boundary conditions (OBC) in a square supercell, as shown in Fig.\ref{fig:corner plot} $(a)$ and $(b)$ for the two representative AM strengths, $\Delta_{\text{AM}}=0.6$ and $0.8$, respectively. We observe the presence of a few in-gap states that are pinned around zero energy and distinctly detached from the bulk (with the probability density, $|\psi|^2$, of those being localized at four corners as shown in the left-up insets) for these AM strengths, clearly indicating a higher-order boundary signature of the OAI phase. For any $\Delta_{\text{AM}}$ in the range $0<\Delta_{\text{AM}}<\Delta_{\text{AM}}^c$, the bulk gap hosts conventional first-order QSH states due to the adiabatic connection of the AM-BHZ model to the original BHZ Hamiltonian, thereby precluding a protected obstruction to realize corner modes. At the critical strength $\Delta_{\text{AM}}^c=0.5$, the altermagnetic $d$-wave splitting becomes large enough to overcome the pre-existing bulk gap, hinting at a topological gap-closing transition. This further alters the crystalline topology of the system for $\Delta_{\text{AM}}>\Delta_{\text{AM}}^c$, where the bulk gap reopens with a modified occupied bandstructure, controlled by $f_{\text{AM}}(k)=(c\text{os}k_x-\text{cos}k_y)$ (protecting AM-symmetry with opposite signs at $X$ and $Y$ momenta under $\mathcal{C}_4$) and $\Delta_{\text{AM}}$ (determining the sign change of the total effective mass of the two adjacent edges) simultaneously, thereby driving the system to an OAI phase with the mass domain-wall corner modes (see details in SM). Focusing on the right-bottom insets of Fig.~\ref{fig:corner plot} we observe that the obstructed phase for $\Delta_{\text{AM}}>\Delta_{\text{AM}}^c$ manifests a quantized fractional corner charge $Q_{\rm cor}\simeq \frac{e}{4}\pmod{e}$ at each corner. This stems from the shifting of the WCCs to an obstructed position, producing a mismatch between electronic charge and boundary termination under OBC, fully consistent with the corner-state accumulation~\cite{Schindler2018,SchindlerII2018,Benalcazar2019}. Therefore, a key finding emerges: a $d$-wave AM embedded in a QSH insulator induces an OAI phase for $\Delta_{\text{AM}}>\Delta_{\text{AM}}^c$ with HOTI-like corner modes.
\begin{figure}
    \centering
    \includegraphics[width=0.8\linewidth]{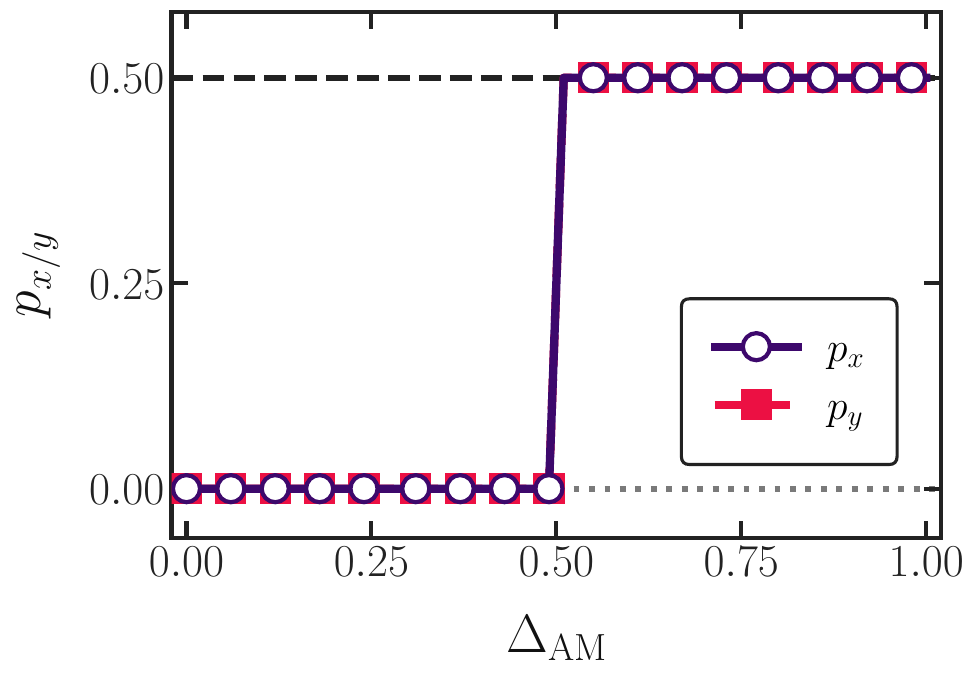}
    \caption{Bulk polarization ($p_{x/y}$) is plotted with $\Delta_{\mathrm{AM}}$. At $\Delta_{\mathrm{AM}}^c=0.5$, both components jump from $p_{x/y}=0$ (which describes a first-order topological phase with $\mathcal{C}_s\neq0$) to the quantized value $p_{x/y}=1/2$, providing a bulk topological characterization of the OAI phase.}
    \label{fig:bulk pol}
\end{figure}
\begin{figure*}
    \centering

    \includegraphics[width=0.95\textwidth]
    {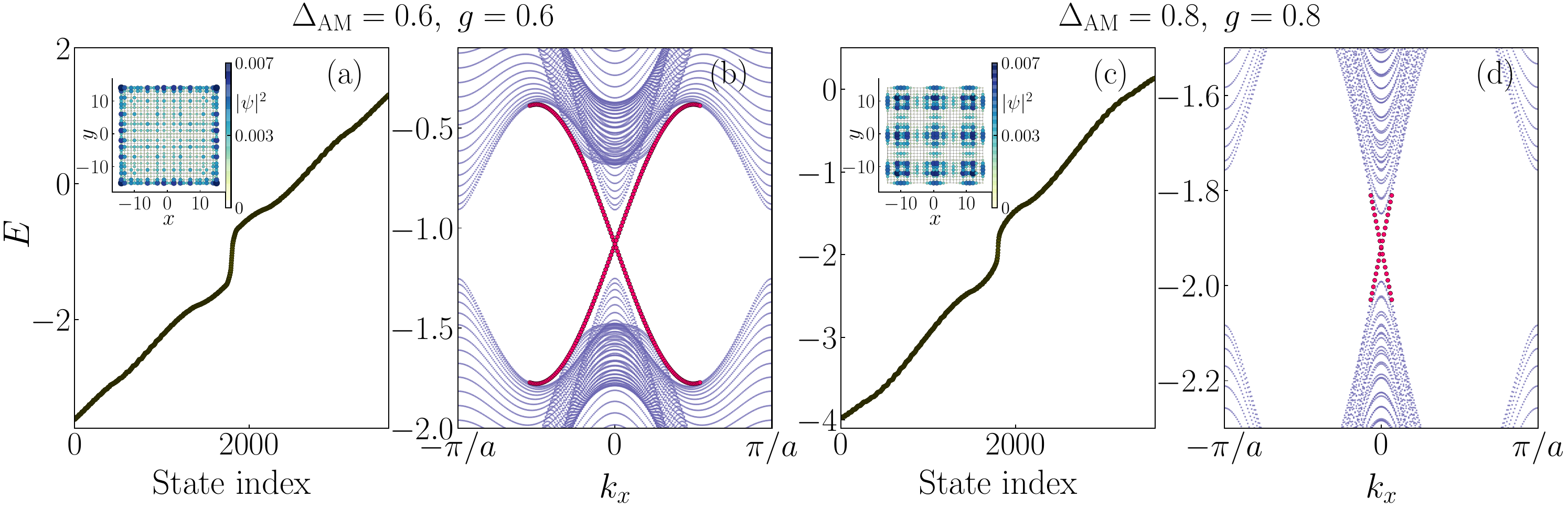}

    \vspace{0.15cm}

    \includegraphics[width=0.95\textwidth]
    {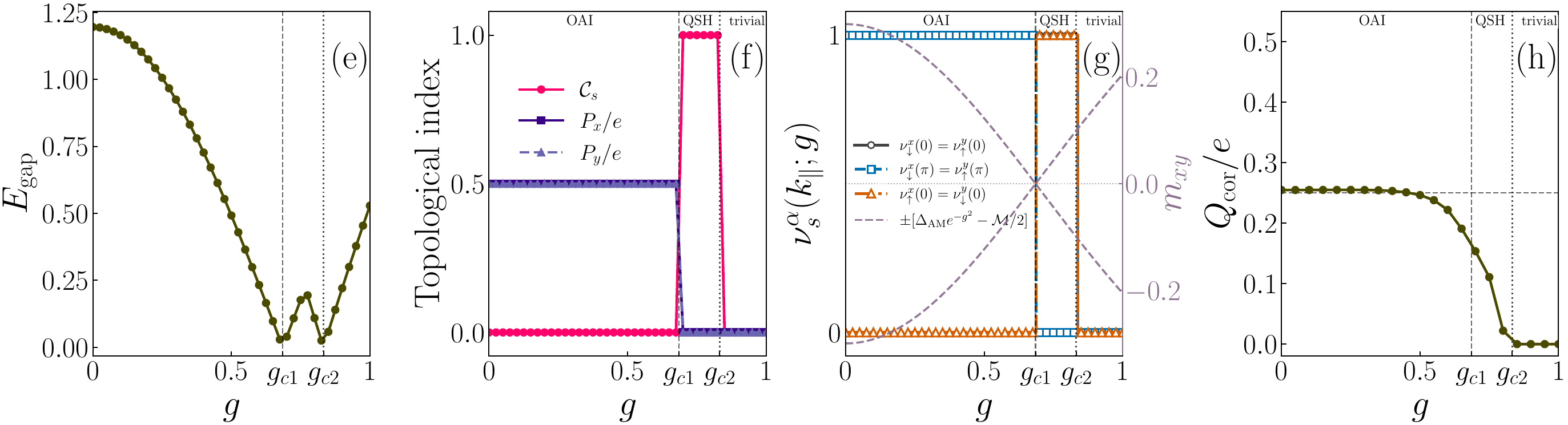}

    \caption{
    Holstein EPC-dressed scenarios: open-boundary energy spectra for $(a)$ $\Delta_{\mathrm{AM}}=0.6$, $g=0.6$ and $(c)$ $\Delta_{\mathrm{AM}}=0.8$, $g=0.8$, with the insets showing the probability density $|\psi|^2$ of the in-gap states, evidencing the vanishing of corner localization of the OAI phase and the restoration of dispersive helical QSH edge modes in corresponding periodic-boundary edge spectra illustrated in $(b)$ and $(d)$, respectively. The lower panel, shown explicitly for $\Delta_{\mathrm{AM}}=0.8$, summarize the phonon-driven evolution with increasing $g$: $(e)$ the bulk gap $E_{\rm gap}$, $(f)$ the topological indices $\mathcal{C}_s$ and $p_{x,y}/e$, $(g)$ the edge-slice invariant $\nu_s^\alpha$ together with the boundary masses $m_{xy}$, and $(h)$ the corner charge $Q_{\mathrm{cor}}$, demonstrating successive OAI-QSH-trivial transitions at critical EPC strengths, $g_{c1}$ and $g_{c2}$, accompanied by a vanishing $Q_{\mathrm{cor}}$ for $g_c>g_{c1}$.
    }
    \label{fig:ep-corner-edge}

\end{figure*}\\
To characterize the altermagnetic OAI phase, we resort to the bulk polarization defined as~\cite{Ezawa2018,Lahiri2024}
\begin{equation}
p_i
=
-\frac{e}{(2\pi)^2}
\sum_{n\in \mathrm{occ}}
\int_{\mathrm{BZ}}
d^2\mathbf{k}\,
\mathrm{Im}
\left\langle
u_{n\mathbf{k}}
\middle|
\partial_{k_i}
u_{n\mathbf{k}}
\right\rangle,~i=x,y.
\end{equation}
The emergence of corner-localized modes in the AM--BHZ model can be understood from the evolution of the occupied WCCs across the altermagnetic transition. In the weak-AM regime, $\Delta_{\mathrm{AM}}<\Delta_{\mathrm{AM}}^c$, the occupied WCCs coincide with the ionic or unit-cell center corresponding to the maximal Wyckoff position $a=(0,0)$, yielding a trivial bulk polarization $\mathbf{p}=(p_x,p_y)=(0,0)$ as shown in Fig.~\ref{fig:bulk pol}.
However, once the AM strength exceeds $\Delta_{\mathrm{AM}}^c$, the term $
H_{\mathrm{AM}}=\Delta_{\mathrm{AM}}\sigma_z\tau_z
$ drives a crystalline-topological reconstruction of the occupied subspace, resulting in a shift of the WCCs away from the ionic center toward an obstructed Wyckoff position~\cite{Bradlyn2017}, $b=\left(\frac12,\frac12\right)$.
Consequently, the bulk polarization becomes quantized~\cite{Benalcazar2017} with $\mathbf{p}=(p_x,p_y)=\left(\frac12,\frac12\right)$,
signalling that the occupied electronic states are displaced by half a lattice vector along both spatial directions (see SM). The robustness of corner localization in the square geometry thus reflects the compatibility between boundary termination and the shifted Wannier configuration, while lattice interactions, such as EPC, may redistribute or weaken the corner response depending on how the boundaries intersect the obstructed Wannier orbitals~\cite{Khalaf2021}.\\
\textit{\textcolor{blue}{Revival of QSH modes by EPC:-}}
The interaction of the phonons with the mobile electrons in an AM-TI can be described by adding a Holstein-type EPC component~\cite{Cangemi2019, Lu2023, Islam2024,Bhattacharyya2024,Lahiri2025} to Eq.~\eqref{Ham:AM BHZ} as
\begin{align}
\mathcal{H}(k)&=
\mathcal{H}_{\text{AM}}^{\text{BHZ}}(k) + \omega_0\mathbb{I}_4\sum_q b_q^\dagger b_q  \nonumber\\&~~~~~~~~~~~~~~~~+\frac{g\omega_0}{\sqrt{N}}\mathbb{I}_4\sum_{k,q}c_{k+q}^\dagger c_k(b_q+b_{-q}^\dagger),
\label{Ham:AM ep-BHZ}
\end{align}
where the vertex of the interaction between longitudinal optical phonon modes oscillating with dispersionless frequency $\omega_0$ and electrons conserves quasi-momentum up to a reciprocal lattice vector $k'=k+q$, with $b_q$ ($b_q^\dagger$) and $c_k$ ($c_k^\dagger$) being the phonon and electron annihilation (creation) operators at momentum $q$ and $k$, respectively. To obtain an effective electronic model we first employ the unitary Lang-Firsov transformation~\cite{Lang1963,Marsusi2012} $\tilde{\mathcal{H}}=\mathcal{U}^{-1}\mathcal{H}\mathcal{U}$ with $\mathcal{U}=\text{exp}[-\frac{g}{\sqrt{N}}\sum_{k,q}c_{k+q}^\dagger c_k(b_q-b_{-q}^\dagger)]$ that describes a shift in the phonon displacements based on the local electronic density in the high frequency anti-adiabatic limit ($\omega_0\gg t,t_{sp},\Delta_{\text{AM}}$), i.e., phononic motion is fast enough to respond rapidly to changes in the electronic configuration, allowing the phononic degrees of freedom to be integrated out to obtain an effective electronic description. The zero-phonon averaging performed on the transformed Hamiltonian ($\tilde{\mathcal{H}}$) yields the coherent phonon-dressed effective electronic model: $\mathcal{H}^{\text{d}}(k)=\langle 0|\tilde{\mathcal{H}}|0\rangle=\sum_kc_k^\dagger\mathcal{H}^{\text{eff}}(k)c_k$ expressed as (see derivation in SM)
\begin{align}{\mathcal{H}}^{\text{eff}}(k)&=
\left[\mathcal{M}-\tilde t(\cos k_x + \cos k_y)\right]\sigma_0\otimes\tau_z + \nonumber \\ &\tilde t_{sp}\left[\sin k_x(\sigma_z\otimes\tau_x)+\sin k_y(\sigma_0\otimes\tau_y)\right]\nonumber\\&+\tilde \Delta_{\text{AM}}(\cos k_x - \cos k_y)\sigma_z\otimes\tau_z -g^2\omega_0\mathbb{I}_4,
\label{Ham:AM_dressed_ep-BHZ}
\end{align}
where the polaronic contributions enter through the reduced hopping strengths denoted by the Holstein band narrowing factor ($\gamma\equiv g^2$) as $(\tilde t, \tilde t_{so}, \tilde{\Delta}_{\text{AM}})=(t,t_{sp},\Delta_{\text{AM}})e^{-\gamma}$ and through the polaronic binding energy $E_p=-g^2\omega_0$, facilitating reconstruction of the bare band topology.
Fig.~\ref{fig:ep-corner-edge} illustrates the evolution of the real-space charge density and energy states of the phonon-dressed AM--BHZ model under OBC for $\Delta_{\mathrm{AM}}=0.6, g=0.6$ ($\Delta_{\mathrm{AM}}=0.8, g=0.8$) as shown in Fig.~\ref{fig:ep-corner-edge}$(a)$ (Fig.~\ref{fig:ep-corner-edge}$(c)$), with the spectrum corresponding to periodic boundaries displayed in Fig.~\ref{fig:ep-corner-edge}$(b)$ (Fig.~\ref{fig:ep-corner-edge}$(d)$). As seen from the real-space charge and spectral distributions, the states rapidly lose their spectral isolation and spatial (corner) confinement upon inclusion of a finite Holstein coupling ($g$), which was previously designated as the AM-induced obstructed phase (Fig.~\ref{fig:corner plot}). Consequently, the associated ribbon spectra develop the characteristic gapless and linearly dispersing (first-order) QSH edges. Physically, the local Holstein EPC suppresses the AM-induced mass reconstruction (through $\Delta_{\text{AM}}\rightarrow\Delta_{\text{AM}}e^{-g^2}$) responsible for the incompatible boundary masses and the obstructed Wannier configuration, leading to the collapse of the corner anomaly and restoration of the gapless QSH edge modes. The lower panels (Figs.~\ref{fig:ep-corner-edge}$(e)-(h)$) for a fixed $\Delta_{\text{AM}}$ clarify the physical mechanism behind the spectral reconstruction for the EPC-induced revival of the QSH boundary in terms of low-energy edge channels (see SM for calculation) expressed as
\begin{equation}
\begin{aligned}
H_{{\rm edge},\eta}^{(\alpha)}(q,g)
&=
\mu(g)\rho_{0}
+
v_{\eta}^{(\alpha)}(g)\,q\,\rho_{z},\\
v_{\eta}^{(\alpha)}(g)
&\propto t_{sp}e^{-g^{2}},
\qquad
\mu(g)=-g^{2}\omega_0 .
\end{aligned}
\end{equation}
where $\alpha=x,y$ denotes the edge orientation and $\eta=0,\pi$ labels the corresponding edge valleys. The topology of each one-dimensional sector is governed by a spin- and momentum-resolved inversion-polarization invariant as~\cite{Hughes2011, Alexandradinata2014} (see SM for details)
\begin{equation}
\begin{split}
\nu_s^{\alpha\in x,y}(k_{\parallel};g)
&=
\frac{
1-\operatorname{sgn}\!\left[
m_s(k_{\perp}=0,k_{\parallel};g)\,
m_s(k_{\perp}=\pi,k_{\parallel};g)
\right]
}{2},
\\
m_s(\mathbf{k},g)
&=
\mathcal{M}-e^{-g^2}[
t(\cos k_x+\cos k_y)
\\
&\hspace{2.5cm}
-s\Delta_{\mathrm{AM}}(\cos k_x-\cos k_y)
].
\end{split}
\label{eq:edge_slice_invariant}
\end{equation}
As $g$ increases, the ubiquitous polaronic factor $e^{-g^{2}}$ progressively weakens the hopping strengths and altermagnetic scales via modifying the relative hierarchy of the Dirac masses at the high-symmetry momenta. The first gap closing (in Fig.~\ref{fig:ep-corner-edge}$(e)$) at $g_{c1}$ occurs when the altermagnetically driven $X/Y$ inversion is removed, while the underlying $\Gamma$-point BHZ inversion remains intact. This converts the obstructed regime, characterized by $p_x=p_y=1/2$ and $\mathcal{C}_s=0$, into a first-order TI with $\mathcal{C}_s=1$ and vanishing bulk polarization, accounting for the re-emergence of the QSH helical boundary modes in the upper spectrum (Fig.~\ref{fig:ep-corner-edge}$(f)$). The edge-resolved slice indices ($\nu_{s}^{x,y}$) in Fig.~\ref{fig:ep-corner-edge}(\textit{g}) further show that this transition redistributes the nontrivial one-dimensional channels, rendering the OAI-specific $k_{\parallel}=\pi$ sector trivial, while restoring the helical $k_{\parallel}=0$ channels characteristic of the QSH phase, concomitant with the sign reversal of the $X/Y$ inversion-control mass $m_{xy}(g)\equiv\pm [\Delta_{\text{AM}}e^{-g^2}-{\mathcal{M}}/2]$ (see SM). Correspondingly, the quantized fractional corner charge in Fig.~\ref{fig:ep-corner-edge}$(h)$ is progressively lost as the corner-localized states delocalize approaching $g_{c1}$. As $g$ increases further, the remaining BHZ band inversion is finally removed at $g_{c2}$, determined by $\Gamma$-point mass \(m_{\Gamma}(g)=[\mathcal{M}-2te^{-g^2}]\), producing the second bulk-gap closing and driving the system into the topologically trivial regime. 
\begin{figure}
    \centering
    \includegraphics[width=0.9\linewidth]{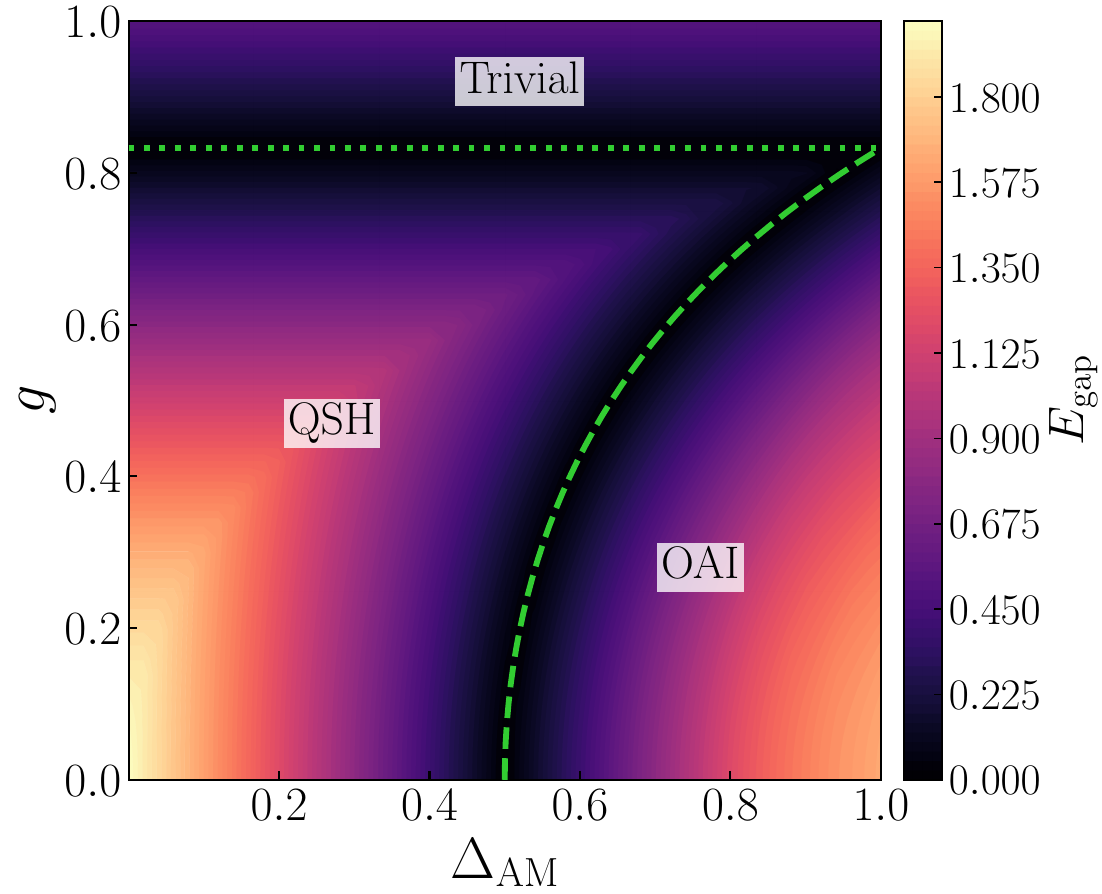}
    \caption{Phase diagram in the $(\Delta_{\rm AM}-g)$ plane for the EPC-dressed AM-BHZ model. The color contour represents the bulk gap $E_{\rm gap}$, while the phase boundaries originating from bulk-gap closings distinguish the OAI, QSH, and trivial insulating regimes, governed by $\mathcal{C}_s$, $p_{x/y}$, and $\nu_{s}^{x,y}$.}
    \label{fig:phase_plot}
\end{figure}\\
\textit{\textcolor{blue}{Conclusion:-}} Taken together, the phase structure summarized in the \((\Delta_{\rm AM}-g)\) plane (Fig.~\ref{fig:phase_plot}) reveals a clear competition between magnetic reconstruction and lattice-induced renormalization in the EPC-dressed AM-BHZ model. The phase boundaries follow from the zeros of the EPC-renormalized masses \(m_{x,y}(g)\) and \(m_{\Gamma}(g)\), with their derivation and the assignment of the intervening gapped regions detailed in the SM. At weak phonon dressing, increasing \(\Delta_{\rm AM}\) drives the parent QSH phase into the OAI regime, characterized by \(p_x=p_y=1/2\), a nontrivial edge-slice pattern, and fractional corner charge. Increasing \(g\), in contrast, reverses this tendency, restoring a window of helical QSH edge transport before the system becomes topologically trivial at stronger EPC. The curved OAI–QSH and nearly horizontal QSH–trivial boundaries highlight the distinct energy scales controlling the two transitions. More broadly, the phase diagram demonstrates how magnetic order and lattice dressing provide complementary control over corner-localized and edge-propagating topological boundary responses within the same microscopic setting.\\
Experimentally, altermagnetic spin splitting has been directly resolved in MnTe by photoemission spectroscopy~\cite{Betancourt2023,Mazin2023,Hariki2024,Krempasky2024}. Meanwhile, the BHZ-QSH phase has been realized in cold-atom experiments~\cite{Lv2021} and in HgTe/CdTe quantum wells, where helical edge transport can be controlled through quantum-well parameters~\cite{Konig2007,Leubner2016}. Importantly, EPC can substantially renormalize Dirac-band inversions and shift topological phase boundaries, with such effects predicted explicitly for HgTe/CdTe quantum wells~\cite{Saha2014}, where EPC strength can be modulated by strain~\cite{Si2016}, surface~\cite{Hwang2013, Aleshkin2024}, electric field~\cite{Yan2007}, or polar molecules (such as CdTe) effects~\cite{Bruderer2007, Herrera2011}. These developments suggest that suitably engineered AM--QSH heterostructures, together with lattice-control techniques, may provide a promising route for probing EPC-controlled transitions between OAI and helical QSH edge modes, where
precise tuning of the system parameters is essential.\\
\textit{\textcolor{blue}{Acknowledgments:-}}
K.B. acknowledges the Anusandhan National Research Foundation (ANRF), Govt. of India, for providing financial support through the National Post Doctoral Fellowship (NPDF) (File No. PDF/2023/000161). K.B. also acknowledges Srijata Lahiri and Debika Debnath for fruitful discussions. \\
\textit{\textcolor{blue}{Data availability statement:-}} The data generated and analyzed in this study are available from the authors
upon reasonable request.
\bibliography{bibfile}


\clearpage
\onecolumngrid

\setcounter{page}{1}
\renewcommand{\thepage}{\roman{page}}

\setcounter{section}{0}
\setcounter{equation}{0}
\setcounter{figure}{0}
\setcounter{table}{0}

\renewcommand{\thesection}{S\arabic{section}}
\renewcommand{\theequation}{S\arabic{equation}}
\renewcommand{\thefigure}{S\arabic{figure}}
\renewcommand{\thetable}{S\arabic{table}}

\begin{center}
{\large\bfseries
Supplemental Material for
``Electron--phonon control of obstructed and quantum Spin Hall
topology in an altermagnet''
}

\vspace{0.5cm}

{\normalsize
Kuntal Bhattacharyya$^{1,*}$,
Bilal Tanatar$^{2}$,
and Saurabh Basu$^{1}$
}

\vspace{0.35cm}

{\small
$^{1}$Department of Physics, Indian Institute of Technology Guwahati,
Guwahati-781039, Assam, India
}

{\small
$^{2}$Department of Physics, Bilkent University,
06800 Bilkent, Ankara, Türkiye
}
\end{center}

\vspace{0.7cm}

\section{Symmetry structure of the $d_{x^{2}-y^{2}}$ altermagnetic BHZ model}
\label{sec:symmetry}

We consider the square-lattice altermagnetic BHZ Hamiltonian with a collinear
$d_{x^{2}-y^{2}}$-wave altermagnet (AM) term as~\cite{Bernevig2006, Hayami2019,Libor2022,LiborII2022,Li-AM2024}
\begin{align}
\mathcal{H}(\mathbf{k})
={}&
\left[
\mathcal{M}-t(\cos k_x+\cos k_y)
\right]\sigma_0\otimes\tau_z
\nonumber\\
&+
t_{sp}
\left[
\sin k_x(\sigma_z\otimes\tau_x)
+
\sin k_y(\sigma_0\otimes\tau_y)
\right]
\nonumber\\
&+
\Delta_{\text{AM}}
(\cos k_x-\cos k_y)
\sigma_z\otimes\tau_z .
\label{eq:S1_fullH}
\end{align}
Here $\mathcal{M}$ is the onsite Dirac mass, $t$ and $t_{sp}$ denote the
intra- and inter-orbital hopping amplitudes, respectively, and
$\Delta_{\text{AM}}$ is the strength of the altermagnetic reconstruction.

\subsection{N\'eel-vector orientation and $d_{x^{2}-y^{2}}$ form factor}

\begin{figure}
    \centering
    \includegraphics[width=1\linewidth]{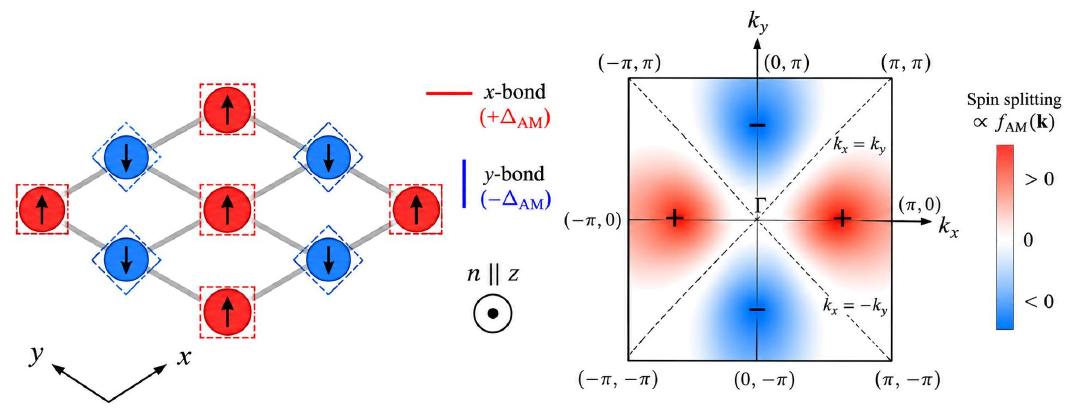}
    \caption{$d_{x^{2}-y^{2}}$ bond anisotropy:
$x$-bonds: $+\Delta_{\rm AM}$, $y$-bonds: $-\Delta_{\rm AM}$,
with Fourier-space form factor
$f_{\rm AM}(\mathbf{k})=\cos k_x-\cos k_y$.
$f_{\rm AM}(\mathbf{k})>0$ for lobes along $\pm k_x$,
$f_{\rm AM}(\mathbf{k})<0$ for lobes along $\pm k_y$, and
$f_{\rm AM}(\mathbf{k})=0$ along the nodal lines $k_x=\pm k_y$.
For the $d_{x^{2}-y^{2}}$ AM-BHZ model,
$\mathcal{C}_4$ and $\mathcal{T}$ are broken, whereas
$\mathcal{P}$, $\mathcal{C}_2$, and $\mathcal{C}_4\mathcal{T}$ are preserved.}
    \label{fig:AM_sym_SM}
\end{figure}

The orientation of the collinear N\'eel order may be parameterized by
\begin{equation}
\hat{\mathbf n}
=
(\sin\theta,0,\cos\theta),
\label{eq:S1_neel}
\end{equation}
so that the altermagnetic contribution can be written as
\begin{equation}
\mathcal{H}_{\rm AM}(\mathbf{k},\theta)
=
\Delta_{\text{AM}} f_{\rm AM}(\mathbf{k})
\left(
\sin\theta\,\sigma_x+\cos\theta\,\sigma_z
\right)\otimes\tau_z ,
\label{eq:S1_AMtheta}
\end{equation}
with
\begin{equation}
f_{\rm AM}(\mathbf{k})
=
\cos k_x-\cos k_y .
\label{eq:S1_formfactor}
\end{equation}
The choice $\theta=0$, adopted in the main text, fixes
$\hat{\mathbf n}\parallel\hat z$ and preserves $s_z$ as a good quantum
number. The $d_{x^{2}-y^{2}}$ designation refers to the crystalline
momentum-space structure of the spin splitting. In real space it follows
from opposite spin-dependent nearest-neighbour reconstructions along the
two principal lattice directions (left panel of Fig.~\ref{fig:AM_sym_SM}),
\begin{equation}
\Delta_x=+\Delta_{\text{AM}},
\qquad
\Delta_y=-\Delta_{\text{AM}},
\label{eq:S1_bondsign}
\end{equation}
whose Fourier transform is proportional to
$\cos k_x-\cos k_y$. The nodal lines of the altermagnetic splitting
therefore occur along $k_x=\pm k_y$ (right panel of Fig.~\ref{fig:AM_sym_SM}).

In the nonrelativistic limit, changing $\theta$ amounts to a global spin
rotation and does not alter the $d_{x^{2}-y^{2}}$ spatial character.
This is distinct from a $d_{x^{2}-z^{2}}$ texture, for which the
crystalline form factor would involve the $x$ and $z$ directions rather
than the $x$ and $y$ directions.
\subsection{Magnetic crystalline symmetries}

For the $d_{x^{2}-y^{2}}$ form factor,
\begin{align}
f_{\rm AM}(-\mathbf{k})
&=f_{\rm AM}(\mathbf{k}),
\\
f_{\rm AM}(\mathcal{C}_{2z}\mathbf{k})
&=f_{\rm AM}(\mathbf{k}),
\\
f_{\rm AM}(\mathcal{C}_{4z}\mathbf{k})
&=-f_{\rm AM}(\mathbf{k}),
\label{eq:S1_ffsym}
\end{align}
where
$\mathcal{C}_{4z}:(k_x,k_y)\mapsto(-k_y,k_x)$.
Time reversal
$\mathcal{T}=i\sigma_y\mathcal{K}$ reverses the spin polarization,
\begin{equation}
\mathcal{T}\sigma_z\mathcal{T}^{-1}
=
-\sigma_z.
\label{eq:S1_Tspin}
\end{equation}
Consequently, the AM term is odd under $\mathcal{T}$ and also odd under
$\mathcal{C}_{4z}$, while the two sign reversals compensate under their
combined operation. Thus inversion $\mathcal{P}$, twofold rotation
$\mathcal{C}_{2z}$, and the antiunitary magnetic crystalline symmetry
$\mathcal{C}_{4z}\mathcal{T}$ remain intact, although $\mathcal{T}$ and
$\mathcal{C}_{4z}$ are individually broken.

\begin{table}[h]
\centering
\caption{
Transformation of the two factors entering
$\mathcal{H}_{\rm AM}=\Delta_{\text{AM}} f_{\rm AM}(\mathbf{k})
\sigma_z\otimes\tau_z$.
}
\label{tab:S1_symmetry}
\begin{tabular}{c c c c}
\toprule
Operation
&
$f_{\rm AM}(\mathbf{k})$
&
$\sigma_z$
&
$\mathcal{H}_{\rm AM}$
\\
\midrule
$\mathcal{P}$ & $+$ & $+$ & $+$ \\
$\mathcal{C}_{2z}$ & $+$ & $+$ & $+$ \\
$\mathcal{T}$ & $+$ & $-$ & $-$ \\
$\mathcal{C}_{4z}$ & $-$ & $+$ & $-$ \\
$\mathcal{C}_{4z}\mathcal{T}$ & $-$ & $-$ & $+$ \\
\bottomrule
\end{tabular}
\end{table}
Since $s_z$ remains conserved for $\theta=0$,
\begin{equation}
\mathcal{H}(\mathbf{k})
=
\mathcal{H}_{\uparrow}(\mathbf{k})
\oplus
\mathcal{H}_{\downarrow}(\mathbf{k}).
\label{eq:S1_block}
\end{equation}
The two $2\times2$ blocks are connected by~\cite{Rafael2025}
\begin{equation}
\mathcal{H}_{\uparrow}(\mathbf{k})
=
\mathcal{U}_{\mathcal{C}_{4}}
\mathcal{H}_{\downarrow}^{*}
(-\mathcal{C}_{4}\mathbf{k})
\mathcal{U}_{\mathcal{C}_{4}}^{\dagger},
\label{eq:S1_C4Tblocks}
\end{equation}
where $\mathcal{U}_{\mathcal{C}_{4}}$ is the orbital representation of
the fourfold rotation. Equation~\eqref{eq:S1_C4Tblocks} is the spin-block
manifestation of the preserved $\mathcal{C}_{4}\mathcal{T}$ symmetry.
The nonrelativistic spin splitting is odd under $x\leftrightarrow y$ and
cancels between the symmetry-related momenta, allowing a momentum-dependent
spin splitting (as shown in Fig.~\ref{fig:AM_sym_SM}) with zero net magnetization.
\section{Wyckoff positions and Wannier-center reconstruction}
\label{sec:wyckoff}

\begin{figure}
    \centering
    \includegraphics[width=1\linewidth]{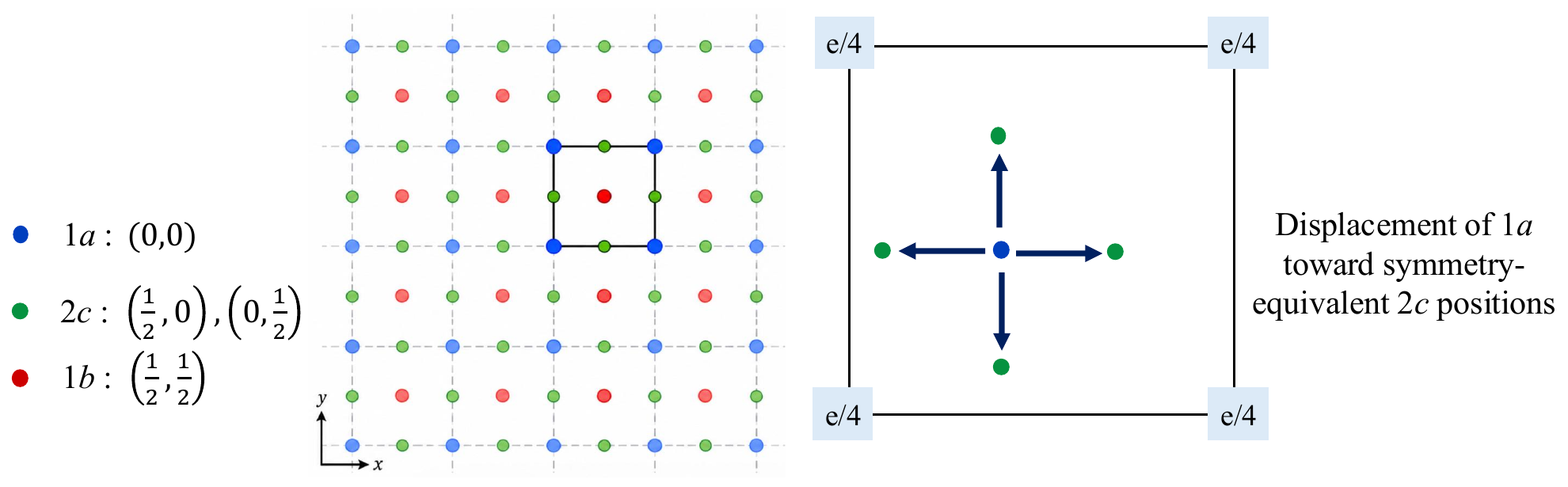}
    \caption{Square-lattice Wyckoff positions:
$1a:(0,0)$ at the lattice site, $1b:(1/2,1/2)$ at the plaquette center, and
$2c:(1/2,0),(0,1/2)$ at the edge centers.
In the OAI regime, the Wannier centers are displaced from the ionic $1a$ reference toward the symmetry-related $2c$ positions, yielding
$\mathbf{p}/e=(1/2,1/2)$ and a fractional corner charge $Q_{\rm cor}=e/4$ under open boundaries.}
    \label{fig:Wyckoff_SM}
\end{figure}

For the square lattice, the maximal Wyckoff positions (as shown in Fig.~\ref{fig:Wyckoff_SM}) relevant to the
present discussion may be represented as
\begin{align}
1a &: (0,0),
\nonumber\\
1b &: \left(\frac12,\frac12\right),
\nonumber\\
2c &: \left\{
\left(\frac12,0\right),
\left(0,\frac12\right)
\right\},
\label{eq:S3_wyckoff}
\end{align}
where the relation between Wyckoff positions and symmetry-compatible
Wannier representations follows the framework of topological quantum
chemistry~\cite{Bradlyn2017,Cano2018}. The physical orbital/ionic site of the minimal BHZ lattice is taken as
the $1a$ reference. A symmetry-preserving displacement of the occupied
Wannier charge centers (WCCs) away from this reference position constitutes the
real-space obstruction.
For exponentially localized occupied Wannier functions with centers
$\mathbf{r}_n$, the electronic polarization satisfies~\cite{King-Smith1993, Vanderbilt1993}
\begin{equation}
\frac{\mathbf{p}}{e}
=
\sum_{n\in{\rm occ}}\mathbf{r}_n
\qquad
(\mathrm{mod}\ \mathbf{R}),
\label{eq:S3_polWCC}
\end{equation}
where $\mathbf{R}$ is a Bravais-lattice vector.
A minimal Wannier representation consistent with the OAI polarization
and the spin-resolved slice pattern is
\begin{equation}
\mathbf{r}_{\downarrow}
\sim
\left(\frac12,0\right),
\qquad
\mathbf{r}_{\uparrow}
\sim
\left(0,\frac12\right),
\label{eq:S3_spinWCC}
\end{equation}
which forms a $2c$-like pair exchanged by
$\mathcal{C}_{4}\mathcal{T}$ (Fig.~\ref{fig:Wyckoff_SM}). Their total polarization is
\begin{equation}
\frac{\mathbf{p}}{e}
=
\left(
\frac12,
\frac12
\right)
\quad
(\mathrm{mod}\ 1),
\label{eq:S3_Phalf}
\end{equation}
consistent with the bulk Wilson-loop result. The nontrivial bulk polarization in the AM-induced regime manifests directly at the boundaries of an open square geometry. Since the polarization is finite along both $x$ and $y$ directions, the edges inherit boundary charge accumulation, and the intersection of two such polarized edges naturally produces enhanced localization near the corners. Microscopically, this can be interpreted as the open boundaries cutting through the shifted Wannier orbitals associated with the obstructed Wyckoff configuration~\cite{Schindler2018,SchindlerII2018,Benalcazar2019}. The resulting corner-localized modes (displayed schematically in the right panel of Fig.~\ref{fig:Wyckoff_SM}) therefore arise as a boundary manifestation of the bulk Wannier obstruction and polarization anomaly. Unlike a ``pure'' higher-order phase characterized by vanishing bulk polarization and a purely secondary corner index, the present AM-induced phase remains polarization active with $p_x=p_y=1/2$, which asserts that the corner response is intrinsically tied to the nontrivial dipole polarization of the occupied bands. In this picture, the corner accumulation may be quantified through the integrated corner charge, $Q_{\mathrm{cor}}
=
\sum_{i\in \mathrm{cor}}
\left(\rho_i-\rho_{\mathrm{bulk}}\right)$, which becomes finite in the obstructed phase. 
Equation~\eqref{eq:S3_spinWCC} should be viewed as the minimal
WCC representation of the numerically obtained polarization
and edge-slice pattern. The precise fractional corner charge is not
inferred from the Wyckoff label alone; it is fixed by the filling anomaly
of the finite geometry~\cite{Benalcazar2019}. Nevertheless, the displacement from the ionic
$1a$ reference to interstitial symmetry-related centers explains why a
square termination leaves uncompensated boundary charge. The transition occurring at $\Delta_{\mathrm{AM}}=\Delta_{\mathrm{AM}}^c$ therefore corresponds not merely to the appearance of localized in-gap states, but to a crystalline-topological reconstruction of the occupied Wannier representation driven by the altermagnetic term.
\section{Derivation of the EPC-dressed altermagnetic BHZ Hamiltonian}
\label{sec:EPC_derivation}

Including the coupling of the dispersionless optical phonon mode with the local electronic density~\cite{Lang1963,Marsusi2012} satisfying~\eqref{eq:S1_fullH}, the total Hamiltonian reads
\begin{align}
\mathcal{H}
={}&
\sum_{\mathbf{k}}
c_{\mathbf{k}}^\dagger
\mathcal{H}_{\rm AM}^{\rm BHZ}(\mathbf{k})
c_{\mathbf{k}}
+
\omega_0
\sum_{\mathbf{q}}
b_{\mathbf{q}}^\dagger b_{\mathbf{q}}
\nonumber\\
&+
\frac{g\omega_0}{\sqrt{N}}
\sum_{\mathbf{k},\mathbf{q}}
c_{\mathbf{k}+\mathbf{q}}^\dagger
c_{\mathbf{k}}
\left(
b_{\mathbf{q}}
+
b_{-\mathbf{q}}^\dagger
\right),
\label{eq:S4_fullH}
\end{align}
where $g$ denotes the local Holstein-type electron--phonon coupling (EPC) strength, $\omega_0$ is the dispersionless
phonon frequency, and $N$ is the number of lattice sites. For notational
simplicity, the identity matrix $\mathbb{I}_4$ in the internal
spin-orbital space is left implicit in the phonon and EPC terms.

To eliminate the linear electron--phonon coupling, we employ the
Lang--Firsov unitary transformation in momentum space as~\cite{Lang1963,Marsusi2012}
\begin{equation}
\widetilde{\mathcal{H}}
=
\mathcal{U}^{-1}
\mathcal{H}
\mathcal{U},
\qquad
\mathcal{U}
=
e^{-\mathcal{R}},
\label{eq:S4_LF}
\end{equation}
with the generator
\begin{equation}
\mathcal{R}
=
\frac{g}{\sqrt{N}}
\sum_{\mathbf{k},\mathbf{q}}
c_{\mathbf{k}+\mathbf{q}}^\dagger
c_{\mathbf{k}}
\left(
b_{\mathbf{q}}
-
b_{-\mathbf{q}}^\dagger
\right).
\label{eq:S4_generator}
\end{equation}
Introducing the Fourier component of the electronic density,
\begin{equation}
\rho_{\mathbf{q}}
=
\sum_{\mathbf{k}}
c_{\mathbf{k}+\mathbf{q}}^\dagger c_{\mathbf{k}},
\label{eq:S4_density}
\end{equation}
the generator can equivalently be expressed as
\begin{equation}
\mathcal{R}
=
\frac{g}{\sqrt{N}}
\sum_{\mathbf{q}}
\rho_{\mathbf{q}}
\left(
b_{\mathbf{q}}-b_{-\mathbf{q}}^\dagger
\right).
\label{eq:S4_generator_density}
\end{equation}

The transformed operators follow from the
Baker--Campbell--Hausdorff expansion,
\begin{equation}
e^{\mathcal{R}}\mathcal{O}e^{-\mathcal{R}}
=
\mathcal{O}
+
[\mathcal{R},\mathcal{O}]
+
\frac{1}{2!}
[\mathcal{R},[\mathcal{R},\mathcal{O}]]
+\cdots .
\label{eq:S4_BCH}
\end{equation}
For the phonon operator one obtains
\begin{equation}
\mathcal{U}^{-1}
b_{\mathbf{q}}
\mathcal{U}
=
b_{\mathbf{q}}
-
\frac{g}{\sqrt{N}}
\rho_{-\mathbf{q}},
\label{eq:S4_btrans}
\end{equation}
and an analogous expression for $b_{\mathbf{q}}^\dagger$. Substitution
into the phonon and interaction sectors cancels the term linear in
$b_{\mathbf{q}}$ and $b_{\mathbf{q}}^\dagger$, yielding
\begin{equation}
\widetilde{\mathcal{H}}_{\rm ph}
+
\widetilde{\mathcal{H}}_{\rm ep}
=
\omega_0
\sum_{\mathbf{q}}
b_{\mathbf{q}}^\dagger b_{\mathbf{q}}
-
\frac{g^2\omega_0}{N}
\sum_{\mathbf{q}}
\rho_{\mathbf{q}}\rho_{-\mathbf{q}} .
\label{eq:S4_ph_ep_trans}
\end{equation}
Within the single-polaron sector relevant to the present effective
single-particle description, the last term reduces to the uniform
polaronic binding-energy shift
\begin{equation}
E_p=g^2\omega_0,
\qquad
\delta\mathcal{H}_{\rm pol}
=
-g^2\omega_0\mathbb{I}_4 .
\label{eq:S4_polaronshift}
\end{equation}

The same transformation dresses the electronic propagation by phonon
displacement operators. The transformed fermionic operator can be
written schematically as
\begin{equation}
\widetilde{c}_{\mathbf{k}}
=
\mathcal{U}^{-1}
c_{\mathbf{k}}
\mathcal{U}
=
\frac{1}{\sqrt{N}}
\sum_{\mathbf{r}}
e^{-i\mathbf{k}\cdot\mathbf{r}}
c_{\mathbf{r}}X_{\mathbf{r}},
\label{eq:S4_ctrans}
\end{equation}
where the displacement cloud is
\begin{equation}
X_{\mathbf{r}}
=
\exp\left[
-\frac{g}{\sqrt{N}}
\sum_{\mathbf{q}}
e^{i\mathbf{q}\cdot\mathbf{r}}
\left(
b_{\mathbf{q}}^\dagger-b_{-\mathbf{q}}
\right)
\right].
\label{eq:S4_Xr}
\end{equation}
Thus, although the derivation is formulated in momentum space, every
momentum-dependent harmonic originating from inter-site electronic
propagation acquires the same phonon-cloud overlap.

In the high-frequency anti-adiabatic regime, the effective coherent
electronic Hamiltonian is obtained by projecting the transformed
Hamiltonian onto the zero-phonon sector (at absolute zero temperature),
\begin{equation}
\mathcal{H}^{d}
=
\langle 0|
\widetilde{\mathcal{H}}
|0\rangle .
\label{eq:S4_zeroav}
\end{equation}
The relevant vacuum expectation value of two phonon displacement clouds
is
\begin{equation}
\langle0|
X_{\mathbf{r}}^\dagger
X_{\mathbf{r}+\boldsymbol{\delta}}
|0\rangle
=
\exp\left(-g^2\right)
\equiv
e^{-\gamma},
\qquad
\gamma=g^2.
\label{eq:S4_Holsteinfactor}
\end{equation}
This is the Holstein band-narrowing factor. Consequently, all terms in
$\mathcal{H}_{\rm AM}^{\rm BHZ}(\mathbf{k})$ originating from
nearest-neighbor propagation are renormalized identically,
\begin{equation}
\widetilde{t}
=
t e^{-\gamma},
\qquad
\widetilde{t}_{sp}
=
t_{sp}e^{-\gamma},
\qquad
\widetilde{\Delta}_{\rm AM}
=
\Delta_{\rm AM}e^{-\gamma}.
\label{eq:S4_dressedparams}
\end{equation}
In particular, $\Delta_{\rm AM}$ acquires the same reduction factor
because the $d_{x^{2}-y^{2}}$ term represents an anisotropic
spin-dependent nearest-neighbor reconstruction, whose Fourier transform
produces the factor $\cos k_x-\cos k_y$. The onsite Dirac mass $M$, on
the other hand, is unaffected by this overlap since it does not involve
electronic propagation between different sites.

Furthermore,
\begin{equation}
\langle0|
b_{\mathbf{q}}^\dagger b_{\mathbf{q}}
|0\rangle
=
0,
\label{eq:S4_zero_phonon}
\end{equation}
so the explicit phonon-number contribution vanishes after the
zero-phonon averaging, while the binding-energy shift in
Eq.~\eqref{eq:S4_polaronshift} remains finite. The resulting effective
momentum-space electronic Hamiltonian is therefore
\begin{align}
\mathcal{H}^{\rm eff}(\mathbf{k})
={}&
\left[
\mathcal{M}
-
\widetilde{t}
(\cos k_x+\cos k_y)
\right]
\sigma_0\otimes\tau_z
\nonumber\\
&+
\widetilde{t}_{sp}
\left[
\sin k_x(\sigma_z\otimes\tau_x)
+
\sin k_y(\sigma_0\otimes\tau_y)
\right]
\nonumber\\
&+
\widetilde{\Delta}_{\rm AM}
(\cos k_x-\cos k_y)
\sigma_z\otimes\tau_z
-
g^2\omega_0\mathbb{I}_4 .
\label{eq:S4_finalHeff}
\end{align}
Equations~\eqref{eq:S4_Holsteinfactor} and
\eqref{eq:S4_dressedparams} show explicitly that the coherent electronic
bandwidth, the inter-orbital hybridization, and the altermagnetic
reconstruction are all reduced by the same zero-phonon overlap, whereas
the spectrum simultaneously acquires the uniform shift
$-g^2\omega_0$~\cite{Cangemi2019,Islam2024,Bhattacharyya2024}. This produces the effective parameters used in the main
text,
\begin{equation}
(\widetilde{t},
\widetilde{t}_{sp},
\widetilde{\Delta}_{\rm AM})
=
(t,t_{sp},\Delta_{\rm AM})
e^{-g^2},
\qquad
E_p=-g^2\omega_0.
\label{eq:S4_summary}
\end{equation}
The different topological sectors therefore arise not from an explicit
phonon population---which vanishes in the zero-phonon projection---but
from the coherent polaronic narrowing of the electronic parameters and
the accompanying binding-energy shift.
\section{Effective edge theory and the origin of the corner modes}
\label{sec:edge}

This section develops the boundary description used to interpret the
AM-induced obstructed phase and its evolution under EPC. A translationally invariant straight edge preserves the
distinction between the high-symmetry $k_{\parallel}=0$ and $\pi$
sectors. Therefore, the appropriate boundary diagnostic is a spin- and
momentum-resolved one-dimensional slice topology rather than an assumed
uniform mass that directly mixes the two valleys.  We emphasize that corner modes do not immediately appear upon introducing the \(d\)-wave altermagnetic term \(\Delta_{\rm AM}f_{\rm AM}(\mathbf{k})\) into the bare BHZ model. Rather, they emerge only after the AM term becomes sufficiently strong to drive the required change in the band inversions and the associated topology.

\subsection{Spin blocks and Dirac masses in the absence of EPC}

For $s=\pm1$ denoting $s_z=\uparrow,\downarrow$, the Hamiltonian reduces
to
\begin{equation}
h_s(\mathbf{k})
=
d_{x,s}(\mathbf{k})\tau_x
+
d_y(\mathbf{k})\tau_y
+
m_s(\mathbf{k})\tau_z ,
\label{eq:S2_spinblock}
\end{equation}
with
\begin{align}
d_{x,s}(\mathbf{k})
&=
s\,t_{sp}\sin k_x,
\\
d_y(\mathbf{k})
&=
t_{sp}\sin k_y,
\\
m_s(\mathbf{k})
&=
\mathcal{M}
-t(\cos k_x+\cos k_y)
+s\Delta_{\text{AM}}(\cos k_x-\cos k_y).
\label{eq:S2_massbare}
\end{align}
At $\Gamma=(0,0)$, $X=(\pi,0)$, $Y=(0,\pi)$, and
$M_{\rm BZ}=(\pi,\pi)\equiv S$,
\begin{align}
m_s(\Gamma)
&=\mathcal{M}-2t,
&
m_s(M_{\rm BZ})
&=\mathcal{M}+2t,
\nonumber\\
m_s(X)
&=\mathcal{M}-2s\Delta_{\text{AM}},
&
m_s(Y)
&=\mathcal{M}+2s\Delta_{\text{AM}} .
\label{eq:S2_TRIMmasses}
\end{align}

\subsection{Spin- and momentum-resolved edge-slice invariant}
For an edge with normal direction \(\alpha=x,y\), \(k_{\parallel}\) remains a good quantum number, while inversion maps \((k_{\perp},k_{\parallel})\rightarrow(-k_{\perp},-k_{\parallel})\); consequently, an individual fixed-\(k_{\parallel}\) slice is inversion symmetric only at \(k_{\parallel}=0,\pi\). Here \(k_{\perp}=k_x\) and \(k_{\parallel}=k_y\) for an \(x\)-normal edge, whereas \(k_{\perp}=k_y\) and \(k_{\parallel}=k_x\) for a \(y\)-normal edge. For each conserved-spin sector \(s\), the corresponding one-dimensional Hamiltonian may be written as \(h_s^{(\alpha)}(k_{\perp},k_{\parallel})=\epsilon_s\tau_0+d_{1,s}\tau_x+d_{2,s}\tau_y+m_s\tau_z\), with inversion represented, up to an overall sign, by \(\mathcal P=\tau_z\). Inversion symmetry requires the coefficients multiplying \(\tau_{x,y}\) to be odd under \(k_{\perp}\rightarrow-k_{\perp}\), whereas \(m_s\) is even; hence, at the one-dimensional inversion-invariant momenta \(K_{\perp}=0,\pi\), the off-diagonal terms vanish and \(h_s^{(\alpha)}(K_{\perp},k_{\parallel})=\epsilon_s(K_{\perp},k_{\parallel})\tau_0+m_s(K_{\perp},k_{\parallel})\tau_z\). The topology of such a one-dimensional inversion-symmetric slice is characterized by the inversion-quantized Zak phase, or equivalently by the quantized electronic polarization along the edge-normal direction, for which
\begin{equation}
    e^{i\gamma_s^{(\alpha)}(k_{\parallel})}
=
\prod_{n\in{\rm occ}}
\xi_{n,s}(0,k_{\parallel})
\xi_{n,s}(\pi,k_{\parallel}),
\qquad
p_{\alpha,s}(k_{\parallel})
=
\frac{e}{2\pi}\gamma_s^{(\alpha)}(k_{\parallel})
\quad (\mathrm{mod}\ e),
\end{equation}
with \(\xi_{n,s}=\pm1\) the occupied-band inversion eigenvalues~\cite{Hughes2011,Alexandradinata2014}. Since each spin block of the present two-band model contains a single occupied band, and since at \(K_{\perp}=0,\pi\) the occupied eigenstate is an eigenstate of \(\tau_z\) with eigenvalue $\xi_s(K_{\perp},k_{\parallel})
=
-\operatorname{sgn}
\!\left[
m_s(K_{\perp},k_{\parallel})
\right],$
one obtains
$$
e^{i\gamma_s^{(\alpha)}(k_{\parallel})}
=
\operatorname{sgn}
\!\left[
m_s(0,k_{\parallel})
m_s(\pi,k_{\parallel})
\right].
$$
Introducing the binary polarization index through
$$
(-1)^{\nu_s^{(\alpha)}(k_{\parallel})}
=
e^{i\gamma_s^{(\alpha)}(k_{\parallel})}
$$ therefore yields a spin-resolved edge-slice invariant as
\begin{equation}
    \nu_s^{(\alpha)}(k_{\parallel})
=
\frac{
1-\operatorname{sgn}\!\left[
m_s(k_{\perp}=0,k_{\parallel})
m_s(k_{\perp}=\pi,k_{\parallel})
\right]
}{2},
\qquad
\alpha=x,y,\quad
k_{\parallel}=0,\pi .
\label{eq:S3_nu}
\end{equation}
Thus, \(\nu_s^{(\alpha)}=1\) is not an independent invariant introduced ad hoc, but the two-band representation of the conventional inversion-quantized one-dimensional polarization: opposite signs of the Dirac mass at \(k_{\perp}=0\) and \(\pi\) imply opposite occupied-state inversion parities, a Zak phase \(\gamma_s^{(\alpha)}=\pi\), and a nontrivial obstructed polarization \(p_{\alpha,s}=e/2\) modulo \(e\), whereas equal mass signs give \(\gamma_s^{(\alpha)}=0\), \(p_{\alpha,s}=0\) modulo \(e\), and \(\nu_s^{(\alpha)}=0\). The invariant is well defined provided the corresponding fixed-spin slice remains gapped. A value $\nu=1$ denotes a topological fixed-spin, fixed-$k_{\parallel}$
slice, whereas $\nu=0$ denotes a trivial one.
\begin{table}[h!]
\centering
\caption{Relation between the Dirac-mass criterion and the conventional inversion-quantized Zak phase and polarization of a fixed-spin, fixed-$k_{\parallel}$ one-dimensional slice.}
\label{tab:mass_invariant_relation}
\begin{tabular}{c c c c c}
\hline\hline
Mass ($m_s(0), m_s(\pi)$) signs &~~~~~
$\xi_s(0)\xi_s(\pi)$ &~~~
$\gamma_s^{(\alpha)}$ &~~~~~
$p_{\alpha,s}$ &~~~
$\nu_s^{(\alpha)}$
\\
\hline
Same: +,+/-,-
& $+1$
& $0$
& $0 \;(\mathrm{mod}\ e)$
& $0$
\\
Opposite: +,-/-,+
& $-1$
& $\pi$
& $\dfrac{e}{2} \;(\mathrm{mod}\ e)$
& $1$
\\
\hline\hline
\end{tabular}
\end{table}

For the AM-induced OAI regime with
$m(\Gamma)<0$, $m(M_{\rm BZ})>0$, and
$\Delta_{\rm AM}>\mathcal{M}/2$, the nontrivial slice invariants for the
$x$- and $y$-normal edges are
\begin{equation}
\begin{aligned}
\mathrm{OAI}:\qquad
&\nu_{\downarrow}^{x}(0)=1,\qquad
\nu_{\downarrow}^{x}(\pi)=1,\qquad
\nu_{\uparrow}^{x}(0)=0,\\
&\nu_{\uparrow}^{y}(0)=1,\qquad
\nu_{\uparrow}^{y}(\pi)=1,\qquad
\nu_{\downarrow}^{y}(0)=0 .
\end{aligned}
\label{eq:S2_OAIpattern}
\end{equation}
while $\mathcal{C}_{4}\mathcal{T}$ gives
\begin{align}
\nu_{\downarrow}^{x}(0)
&=
\nu_{\uparrow}^{y}(0),~~~
\nu_{\downarrow}^{x}(\pi)
=
\nu_{\uparrow}^{y}(\pi),
\nonumber\\
\nu_{\uparrow}^{x}(0)
&=
\nu_{\downarrow}^{y}(0),~~~
\nu_{\uparrow}^{x}(\pi)
=
\nu_{\downarrow}^{y}(\pi).
\label{eq:S2_C4Tnu}
\end{align}
\subsection{Straight-edge theory and corner localization}

Expanding around a boundary valley $k_{\parallel}=\eta+q$,
$\eta=0,\pi$, a low-energy edge channel can be represented as
\begin{equation}
H_{{\rm edge},\eta}^{(\alpha)}(q)
=
v_{\eta}^{(\alpha)}q\,\rho_z,
\qquad
\alpha=x,y,
\label{eq:S2_edgebare}
\end{equation}
where $\rho_i$ acts in the projected low-energy boundary space.
The existence of a branch is controlled by
Eqs.~\eqref{eq:S3_nu}.

On an ideal straight edge, translation symmetry prevents local mixing between the $k_{\parallel}=0$ and $\pi$ boundary valleys, since such a process requires a momentum transfer of $\pi$. At a corner, however, $k_{\parallel}$ is no longer conserved, and the corresponding intervalley coupling becomes symmetry allowed. Upon projection onto the two boundary valleys, this coupling acts as an effective Dirac mass, leading to the standard mass-domain-wall description of a higher-order corner state~\cite{JackiwRebbi1976,Langbehn2017,Khalaf2018,Schindler2020}.
Near the corner the minimal local theory is~\cite{Schindler2020}
\begin{equation}
H_{\rm cor}
=
-i v_{\rm e}\rho_z\partial_{\ell}
+
\Delta_{\rm iv}(\ell)\rho_x ,
\label{eq:S2_cornerDirac}
\end{equation}
where $\ell$ parametrizes the boundary through the corner.
If the effective intervalley coupling interpolates between
symmetry-related terminations with opposite signs, the associated Jackiw--Rebbi
domain wall~\cite{JackiwRebbi1976,Schindler2020}
supports a localized zero mode,
\begin{equation}
\psi_{\rm cor}(\ell)
\propto
\exp\left[
-\int_{0}^{\ell}
\frac{\Delta_{\rm iv}(\ell')}{v_{\rm e}}
\,d\ell'
\right]\chi .
\label{eq:S2_cornerwf}
\end{equation}
Thus, the straight edges need not carry a uniform valley-mixing mass,
whereas the corner supplies the translation-breaking process required to
localize the OAI boundary spectral weight.

The same result has a filling-anomaly interpretation. In the OAI regime
the occupied charge centers are displaced from the ionic reference
positions, so an open square cannot terminate all Wannier centers
locally while maintaining the bulk filling. In the finite geometry used
in the main text, half filling of the near-midgap sector distributes one
electronic charge among the four symmetry-related corners~\cite{Benalcazar2017},
\begin{equation}
Q_{\rm cor}
=
\frac{e}{4}
\qquad
(\mathrm{mod}\ e).
\label{eq:S2_cornercharge}
\end{equation}

\subsection{EPC-dressed edge theory and restoration of the QSH phase}

The Lang--Firsov procedure derived in Sec.~\ref{sec:EPC_derivation} yields
\begin{equation}
t\rightarrow t e^{-g^2},
\qquad
t_{sp}\rightarrow t_{sp}e^{-g^2},
\qquad
\Delta_{\text{AM}}\rightarrow \Delta_{\text{AM}} e^{-g^2},
\label{eq:S2_dressedparams}
\end{equation}
together with
\begin{equation}
\mu(g)=-g^2\omega_0.
\label{eq:S2_mu}
\end{equation}
The dressed mass is
\begin{equation}
\begin{aligned}
m_s(\mathbf{k},g)
={}&
\mathcal{M}
-t e^{-g^2}
(\cos k_x+\cos k_y)
\\
&+
s\Delta_{\text{AM}} e^{-g^2}
(\cos k_x-\cos k_y),
\end{aligned}
\label{eq:S2_massg}
\end{equation}
and the low-energy edge Hamiltonian becomes
\begin{equation}
\begin{aligned}
H_{{\rm edge},\eta}^{(\alpha)}(q,g)
&=
\mu(g)\rho_0
+
v_{\eta}^{(\alpha)}(g)q\,\rho_z,
\\
v_{\eta}^{(\alpha)}(g)
&\propto
t_{sp}e^{-g^2}.
\end{aligned}
\label{eq:S2_edgeg}
\end{equation}\\
Since $t_{sp}\neq 0$, the momentum-odd terms proportional to
$\sin k_x$ and $\sin k_y$ vanish simultaneously only at the
time-reversal-invariant momenta, so that a bulk-gap closing can occur
only when the corresponding Dirac mass $m_s(\mathbf{k},g)$ vanishes
at one of these momenta. In particular, at \(X/Y\), the spin-resolved Dirac masses are \(m_s(X/Y,g)=\mathcal M\mp2s\Delta_{\rm AM}e^{-g^2}\). Equivalently, the analytical gap-closing condition is captured by the effective masses \(m_{x,y}(g)\equiv\pm[\Delta_{\rm AM}e^{-g^2}-\mathcal M/2]\) introduced in the main text, whereas the gap at \(\Gamma\) is governed by \(m_{\Gamma}(g)=[\mathcal{M}-2te^{-g^2}]\). Their zeros yield the two transition conditions below. The first EPC-driven transition occurs when
\begin{equation}
\mathcal{M}
-2\Delta_{\rm AM}e^{-g_{c1}^{2}}=0,
\qquad
g_{c1}
=
\sqrt{
\ln\!\left(
\frac{2\Delta_{\rm AM}}{\mathcal{M}}
\right)
},
\label{eq:S2_gc1}
\end{equation}
provided $2\Delta_{\rm AM}>\mathcal{M}$. Equivalently, solving the first gap-closing condition for the altermagnetic strength yields
$\Delta_{\rm AM}^{c}(g)=\frac{\mathcal{M}}{2}e^{g^{2}}$,
which defines the OAI–QSH phase boundary in the \((\Delta_{\rm AM}-g)\) plane and corresponds to the curved dashed line in the phase diagram of the main text.
Above this transition, while the BHZ inversion is still present, the
slice pattern becomes
\begin{equation}
\begin{aligned}
\mathrm{QSH}:\qquad
&\nu_{\downarrow}^{x}(0)=1,\qquad
\nu_{\uparrow}^{x}(0)=1,\qquad
\nu_{\downarrow}^{x}(\pi)=\nu_{\uparrow}^{x}(\pi)=0,\\
&\nu_{\downarrow}^{y}(0)=1,\qquad
\nu_{\uparrow}^{y}(0)=1,\qquad
\nu_{\downarrow}^{y}(\pi)=\nu_{\uparrow}^{y}(\pi)=0 .
\end{aligned}
\label{eq:S2_QSHpattern}
\end{equation}
The OAI-specific $\pi$ sector disappears and the two
$k_{\parallel}=0$ spin channels form the restored helical QSH boundary.
Accordingly, the isolated corner sector is no longer supported, and its spectral weight evolves into the extended helical edge response.

The second transition occurs when
\begin{equation}
\mathcal{M}-2t e^{-g_{c2}^2}=0,
\qquad
g_{c2}
=
\sqrt{
\ln\left(\frac{2t}{\mathcal{M}}\right)
},
\label{eq:S2_gc2}
\end{equation}
provided $2t>\mathcal{M}$. Beyond $g_{c2}$,
\begin{equation}
\text{trivial:}\qquad
\nu_s^{x}(0)
=
\nu_s^{x}(\pi)
=
\nu_s^{y}(0)
=
\nu_s^{y}(\pi)
=
0.
\label{eq:S2_trivialpattern}
\end{equation}
Combining the conditions for the two transitions, an intermediate EPC-restored QSH regime occurs for
$\mathcal{M}/2 < \Delta_{\rm AM} < t$, with $g_{c1} < g < g_{c2}$; at
$\Delta_{\rm AM}=t$, the two critical couplings coincide and the intermediate QSH window closes. Since the relevant symmetries are preserved throughout the parameter sweep, the topological indices remain unchanged within each gapped region and can change only across a bulk-gap closing. The phase diagram in the \((\Delta_{\rm AM}-g)\) plane in the main text is therefore assigned by continuity from the representative OAI, QSH, and trivial regimes characterized above through $\nu_s^{(\alpha)}$, \(C_s\), and \(p_{x/y}\), with the phase boundaries determined by the analytical gap-closing conditions \eqref{eq:S2_gc1} and \eqref{eq:S2_gc2}. The three regimes can therefore be distinguished in real space as
follows. In the OAI, the occupied WCCs are displaced from the
ionic reference, $\mathbf{p}/e=(1/2,1/2)$, and the boundary carries the
nontrivial $0/\pi$ slice pattern. In the QSH phase the obstructed
polarization is removed, but the occupied subspace retains a nontrivial
first-order QSH topology and therefore cannot be reduced to the same
atomic limit while preserving the relevant topology. In the trivial
regime, the first-order topology is also lost and the WCCs can
be continuously returned to the atomic reference configuration, with
$\mathbf{p}=0$ and all edge-slice indices trivial.

Importantly, this mechanism does not require the specific exponential
renormalization produced by the Holstein--Lang--Firsov treatment.
More generally, the restoration of the QSH phase only requires an
EPC-induced renormalization for which the altermagnetic $X/Y$ inversion
is removed before the underlying BHZ inversion at $\Gamma$. Thus, a
common multiplicative narrowing of the intersite terms, as realized
here by $e^{-g^2}$, is sufficient but not necessary for the transition.
If a more general EPC renormalizes the hopping, altermagnetic term, and
local mass differently, the same mechanism persists provided this
ordering of the two band inversions is maintained. The corresponding
critical couplings and the width of the intermediate QSH regime will,
however, become model dependent.


\end{document}